# Orthogonal Superposition Rheometry of soft core-shell microgels

*Panagiota Bogri[1,2], Gabriele Pagani[3], Jan Vermant[3], Joris Sprakel[4] and George Petekidis[1,2]*

1. Department of Material Science and Technology, University of Crete, Heraklion, Greece

2. IESL/FORTH Heraklion, Crete, Greece

3. Department of Materials, ETH Zurich, CH-8093 Zurich, Switzerland

4. Department of Agrotechnology and Food Sciences, University of Wageningen, The Netherlands

**Abstract:**

The mechanisms of flow in suspensions of soft particles above the glass-transition volume fraction and in the jammed state were probed using Orthogonal Superposition Rheometry (OSR). A small amplitude oscillatory shear flow is superimposed orthogonally onto a steady shear flow, which allows monitoring the viscoelastic spectra of sheared jammed core-shell microgels during flow. The characteristic crossover frequency $\omega_c$, deduced from the viscoelastic spectrum, provides information about the shear induced structural relaxation time, which is connected to the microscopic yielding mechanism of cage breaking. The shear rate evolution of the crossover frequency is used to achieve a superposition of all spectra and get a better insight of the flow mechanism. Despite their inherent softness, the hybrid core-shell microgels exhibit similarities with hard sphere-like flow behavior, with the main difference that for the microgels the transition from a glassy to a jammed state introduces a volume fraction dependence of the scaling of $\omega_c$ with shear rate. We further check the application of the Kramers-Kronig relations on the experimental low strain amplitude OSR data finding a good agreement. Finally, the low frequency response at high strain rates was investigated with open bottom cell geometry and instrumental limits were identified. Based on these limits, we discuss previous OSR data and findings in repulsive and attractive colloidal glasses, and compared them with the current soft particle gels.

## I. Introduction:

The study of the rheological behavior of complex materials under flow and its interplay with the microstructure, consist of an active research topic in soft matter physics that has received much attention (De Silva *et al.* 2011, Mewis and Wagner 2011, Eberle and Porcar 2012, Koumakis *et al.* 2012b, Lettinga *et al.* 2012, López-Barrón *et al.* 2012, Koumakis *et al.* 2013, Amann *et al.* 2015, Snijkers *et al.* 2015, Vlassopoulos and Cloitre 2021, Wagner and Mewis 2021). Colloidal hard spheres have been extensively used as model systems, to study phase behavior, state transitions and flow response of both ordered and glassy states (Koumakis *et al.* 2008, Koumakis *et al.* 2012a). For hard spheres the glass transition occurs at volume fractions above $\phi_g \approx 0.58$, due to entropic caging of particles induced by the presence of their first neighbors (Pusey 1991, van Megen and Underwood 1994, Cheng *et al.* 2001, El Masri *et al.* 2009). The mechanisms of yielding and flow of hard sphere colloidal glasses have been studied through experiments and simulations, for both steady (Petekidis *et al.* 2004, Schall *et al.* 2007, Besseling *et al.* 2010, Chikkadi *et al.* 2011, Koumakis et al. 2012a, Siebenbürger *et al.* 2012) and oscillatory (Petekidis *et al.* 2002, Miyazaki *et al.* 2006, Brader *et al.* 2010, Koumakis et al. 2013) shear flow. Furthermore, Jacob *et al.* (2015) studied the viscoelastic spectra of hard sphere glasses under steady shear flow and the shear rate dependence of structural relaxation using Orthogonal Superposition Rheometry, the experimental technique reported in this investigation.

Similar to hard spheres, the mechanism of yielding and flow has also been studied for soft systems, such as microgels and star-like systems (polymer grafted colloids and multiarm stars) (Cloitre *et al.* 2003, Erwin *et al.* 2010, Seth *et al.* 2011, Truzzolillo *et al.* 2014, Vlassopoulos and Cloitre 2014, 2021). Soft systems undergo the same qualitative sequence of the phase transitions with hard sphere suspensions, with the difference that, due to the deformability and compressibility of the microgels and the interpenetrability of star-like systems, the glass and the jamming transitions are shifted to higher volume fractions, with the possibility to be larger than one (Mewis and Wagner 2011, Lyon and Fernandez-Nieves 2012, Vlassopoulos and Cloitre 2014, 2021). In addition, non-linear rheology in connection with microstructural evolution has been explored in several soft glasses under shear (Le Grand and Petekidis 2008, Siebenbürger *et al.* 2009, Mohan *et al.* 2013).

In this work, we explore the effect of steady shear on the viscoelastic behavior of sheared hybrid core-shell microgels well above glass transition through Orthogonal Superposition Rheometry (OSR). OSR combines two deformation modes, steady shear in one direction and small amplitude oscillatory shear applied simultaneously and orthogonally to the steady shear. In this way, small amplitude orthogonal frequency sweeps interrogate the sample which is subjected to steady shear flow enabling the direct measurement of its viscoelastic spectra (Simmons 1966, Vermant *et al.* 1998, Kim *et al.* 2013, Jacob et al. 2015).

The OSR approach has been so far utilized in a number of soft matter systems including polymer solutions and colloidal gels, dilute or concentrates as well as colloidal glasses of soft and hard sphere particles, undergoing steady shear (Mewis *et al.* 2001, Mobuchon *et al.* 2009, Kim et al. 2013, Jacob et al. 2015, Khandavalli *et al.* 2016, Colombo *et al.* 2017, Sung *et al.* 2018, Moghimi *et al.* 2019, Rathinaraj *et al.* 2022, Pagani *et al.* 2024). In some of these studies of colloidal glasses and gels we came across a common phenomenology, although clear system specific differences have also been revealed. With regard to the first an interesting, yet still puzzling, observation is a strengthening of the elastic response (G'>G'') observed at low frequencies at rather high steady shear rates. This finding has been attributed to possible precursors of shear thickening effects, which however are yet to be verified. Here we revisit this point using a similar type of soft colloidal gel to interrogate the underlying viscoelastic spectra under steady shear, with emphasis on the low frequency regime. For this purpose, we utilize a new type of Couette shear cell custom made for OSR measurements to eliminate pumping effects that might ruin delicate measurements in yield stress samples at low frequencies. Careful analysis at the low frequency regime with all different rheological tools available applied to the soft glassy sample reveals the instrumental limit at low frequency. This suggests that large part of the earlier reported low frequency elasticity was due to the instrumental pumping effects. We finally look into the character of measured viscoelastic spectra under steady shear and confront it with the Kramers-Kronig relations, which follow from the causality condition, but may be expected for some systems not to hold in non-conservative flows such as shearing, at least under certain conditions (Dhont and Wagner 2001).

II. Experimental

*Systems*

We used temperature-sensitive soft core-shell particles consisting of a fluorescently labeled (Nile red) trifluorethyl methacrylate (TFEMA) core of 85nm radius onto which a Poly(*N*-isopropylacrylamide) shell is affixed. The composition of the microgel shell determines the thermosensitive behavior of the hybrid particle. The hybrid particles, TFEMA/PNIPAM (diameter 870 nm), are synthesized in a two step-procedure according to the process described by Appel *et al.* (2015). The crosslinking density of the shell is 5%wt. The particles which are electrostatically stabilized in aqueous suspensions by sulfonate groups at their periphery, have a hydrodynamic radius of $R_H$=435 nm at T=20 $^{\circ}$C as determined by dynamic light scattering in the dilute regime. The effective volume fraction $\phi_{eff}$ of colloidal suspensions was extracted by measuring the relative viscosity $\eta_{rel}$ of dilute suspensions as a function of particle concentration, using capillary viscometry, and subsequent fitting of the volume fraction dependence (in the dilute regime) by the Batchelor-Einstein prediction for hard spheres.

*Rheometry*

Orthogonal Superposition Rheometry is performed at various rotational rates and three different effective volume fractions well above the glass transition (0.99, 1.13 and 1.27 at T=20 $^{\circ}$C), using an ARES-G2 (TA Instruments) strain-controlled rheometer, equipped with a modified double wall Couette geometry (Vermant *et al.* 1997, Kim et al. 2013). The control loop of the normal force transducer is adapted to impose a small strain amplitude oscillatory motion orthogonal to the standard rotational motion of the rheometer (Vermant et al. 1997, Kim et al. 2013). Orthogonal Dynamic frequency sweeps were performed at a low strain amplitude in linear regime (γ=0.3%) once a steady flow was reached. We explored tangential steady shear rates, $\dot{\gamma}$, ranging from $10^{-3}$ to 10 s$^{-1}$. For this range of shear rates, no measurable wall slip was revealed according to the flow curves in Figure 1, where a well-defined yield stress plateau was measured indicative of non-slip conditions (Ballesta *et al.* 2008, Ballesta *et al.* 2013). All experiments were performed in the repulsive regime (with T < $T_{LCS}$ ~ 34 °C) at T=20 °C.

In addition, a special Open Bottom Couette (OBC) geometry was used to verify the effect of pressure gradients. The radii of the cell are $R_i$=5 mm (outer) and $R_o$=6 mm (inner) and the measuring length is L=25 mm (see schematic in figure A1). This prototype cell was mounted on a linear motor for an MCR702 (Anton-Paar) stress-controlled rheometer. The yield stress of the material keeps the sample in the gap between the concentric cylinders without dripping of the material into the open volume at the bottom of the bob. In this way this geometry minimizes pressure effects generated by the displacement of the yield-stress material in the material reservoir, typical in the normal OSP Couette. The ability of yield stress material to sustain their own weight for short period of time is also important in the loading procedure. The cavity was loaded with an excess of the material to ensure complete filling. Then, the internal geometry was removed and the excess material in the bottom cavity cleaned. The lower edge was trimmed before reinserting the bob and the loading procedure was concluded with the trimming of the upper edge. The described procedure ensures that the material fills the measuring annulus effectively. It should be noted that these are the first data acquired with this new prototype cell/geometry. However, since the geometry is not covered by a humidity controlling cover that would minimize solvent evaporation the material was subject to some evaporation and thus measurements were limited to short times. Note that the effective volume fraction for these measurements was estimated to be $\phi_{eff}$ =1.37, i.e. higher than the ones used with the normal OSR Couette, mentioned above. The higher effective volume fraction ensures that the measurements fall into the operational window of the instrument. The data acquisition procedure and the corrections applied are described in the Appendix.

### III. Results:

First, we present the investigation of the flow behavior of the system under constant steady shear. Figure 1 shows the flow curves (steady state shear stress versus shear rate) for various volume fractions. These curves present the typical response expected for soft yield-stress materials with a shear thinning behavior at high shear rates, where stress shows a sub-linear increase with shear rate and the detection of a yield stress plateau at low shear rates. The shear stress and viscosity are shown in figure 1 as a function of the dimensionless Péclet number ($\text{Pe} = \dot{\gamma} t_B = \dot{\gamma} R^2/D_O$ with $t_B = 0.384$ s being the free Brownian time of the particle with R=870 nm at T=20 °C, with $D_O$ is the diffusivity in the dilute regime calculated through the Stokes-Einstein-Sutherland relation), which relates the shear rate of a flow to the particle's diffusion rate. The shear stress can be fitted with the Herschel-Bulkley model, $\sigma = \sigma_0 + k\dot{\gamma}^n$, in agreement with similar earlier studies (Cloitre et al. 2003, Bécu *et al.* 2006, Ballesta et al. 2008, Erwin et al. 2010, Koumakis et al. 2012b). Table I displays the values of k and n for various volume fractions. The consistency factor, k, is a constant of proportionality, that increases with increasing volume fraction. The flow index exponent, n, describes the shear-thinning behavior and shows an insignificant change with volume fraction with value less than one in the high shear rate regime measured.

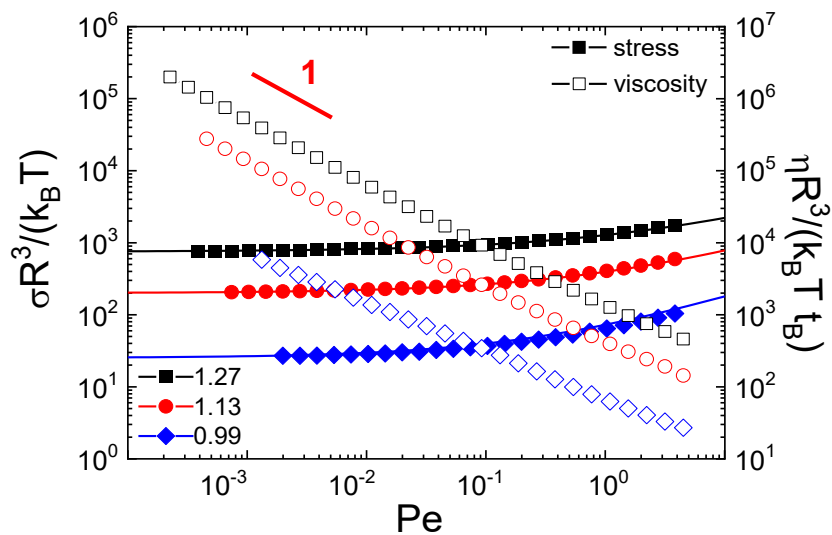

*Figure 1. Normalized steady state shear stress (closed symbols, left axis) and shear viscosity (open symbols, right axis) as a function of scaled shear rate (Pe) for various volume fractions at T=20 °C. The unscaled data are shown in Figure S1 of the supplementary material.*

| $\phi_{eff}$ | k | n |
|---|---|---|
| 1.27 | 523 | 0.44 |
| 1.13 | 195 | 0.48 |
| 0.99 | 47 | 0.51 |

*Table I - The consistency factor, k, and shear thinning exponent, n, values obtained by fitting experimental data of Figure 1 with the Herschel-Bulkley model.*

The extent of the linear viscoelastic regime in the orthogonal direction is determined by performing dynamic strain sweep (DSS) tests. Figure 2a depicts the orthogonal dynamic strain sweep for $\phi_{eff}$=1.27 at rest as well as for rotational shear rates of 0.1 and 1 s$^{-1}$ at the fixed frequency of 10 rad/s. The rotational dynamic strain sweep is also depicted for comparison. The rotational DSS data exhibit the typical features for yielding as seen in many colloidal glasses and jammed states of soft particles (Petekidis et al. 2002, Koumakis et al. 2012a, Koumakis et al. 2012b), i.e. a decrease of G' with increasing strain amplitude and the simultaneous appearance of a peak in G'' at approximately 10% strain amplitude, as well as the constant power law slopes at high strains with G' having roughly twice the slope of G''. With no shear, the orthogonal dynamic strain sweep exhibits similar response to the parallel one with a small discrepancy in the plateau value. The difference can be caused by end-effects or anisotropy induced by the pre-shear. As expected, the onset of non-linearity begins at around 1% strain, with higher volume fraction showing an earlier onset. Increasing the shear rate, the system tends to liquefy leading to the extension of the linear regime to larger strain amplitude values. For $\phi_{eff}$=1.27 the linear regime at rest extends to approximately 0.3%, whereas under steady shear rate of 0.1 s$^{-1}$ it increases to 1.5% and at 1 s$^{-1}$ to 3%. In dynamic strain sweeps, the peak of loss moduli, G'', represents the maximum energy dissipation around yielding where it is expected that a large number of cages break and particle rearrangements are important (Petekidis et al. 2002, Koumakis et al. 2012a). Here, we observe that under steady shear at increasing shear rate the peak disappears indicating a more gradual elastic energy dissipation during yielding through microscopic rearrangement. During steady state shear flow, there is plastic deformation, which continuously relaxes the elastic strain. Therefore, while the LAOS G'' peak at rest indicates a transition from solid to liquid response, accompanied by a peak in energy dissipation, for the flowing sample under steady shear this transition has already taken place. Therefore, the OSR LAOS simply probes the residual structural response that transits from probing restructuring cages under low strain amplitudes to stronger shear thinning response due to additional deformation of the structure when the orthogonal strain amplitude increases.

Figure 2b presents the orthogonal dynamic frequency sweeps for $\phi_{eff}=1.27$, at a low strain amplitude in linear regime, which were performed once a steady state was reached, typically after 200 s for the highest to 2000 s for the lowest shear rates. The waiting time used each time corresponds to a total strain of γ>1. The dynamic frequency sweeps are represented both as a function of the angular frequency ω or the dimensionless frequency $Pe_\omega = \omega t_B$, the latter differ from the Péclet number by the absence of any information regarding the amplitude of the deformation. The orthogonal DFS experiment at quiescent conditions shows the typical behaviour of colloidal suspensions above glass transition where the storage modulus G' dominates the loss modulus G'' over a wide range of frequencies. The storage modulus G' presents a plateau with a slightly increasing slope, which is due to increased particle motions at low frequencies compared to high ones, where the motion is restricted due to the caging effect. The loss modulus G'' exhibits a minimum, which is related to the transition between the in-cage and the out-of-cage relaxation times of the particles (van Megen and Underwood 1993, Mason and Weitz 1995, Koumakis et al. 2012b). Moreover, the evolution of the orthogonal elastic, $G'_\perp$, and viscous, $G''_\perp$, moduli at various rotational shear rates is presented, where upon increasing the main shear rate, the stress induces stronger microstructural changes in the system, which results in speeding up of the internal relaxation dynamics.

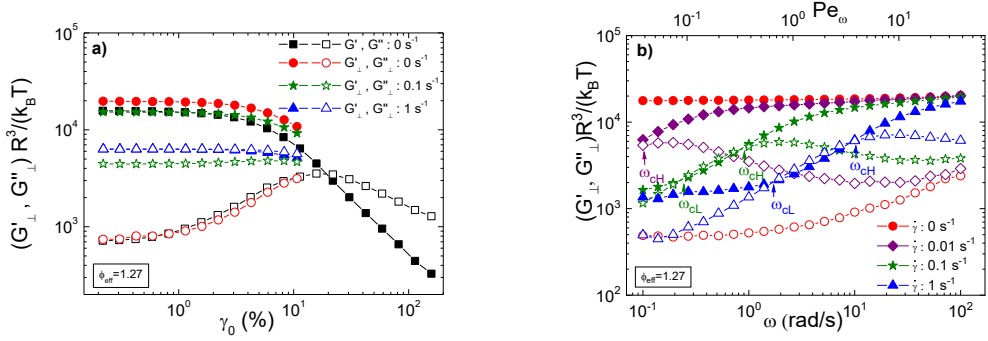

*Figure 2. a) Orthogonal Dynamic Strain Sweeps (ODSS) showing the elastic, $G'_\perp$, (solid symbols) and viscous, $G''_\perp$, (open symbols) moduli at frequency of ω=10 rad/s under steady shear flow for various shear rates as indicated at $\varphi_{eff}$ =1.27 of TFEMA/PNIPAM 870nm. The parallel dynamic strain sweeps (black curves) are also shown for the comparison; b) Storage and loss moduli from the orthogonal frequency sweeps at $\varphi_{eff}$ =1.27 at T=20 °C for various shear rates as a function of frequency, ω (bottom x-axis), and the scaled frequency (top x-axis), $Pe_\omega = \omega t_B$*

From the dynamic frequency sweeps we can extract the plateau modulus $G'_p$, located at the minimum of G'' around a frequency $\omega_m$ (Pellet and Cloitre 2016). Figure 3 shows the plateau modulus as a function of effective particle volume fraction. For the TFEMA/PNIPAM (870 nm) system, although we do not have a lot of data points, we clearly observe a change of slope with increasing the effective volume fraction. This change in concentration dependence of the plateau modulus, taking place around $\varphi_{eff}$ =1, may signify a transition from glassy to jammed state as has been reported by Pellet and Cloitre (2016) as a mechanical signature of glass to jammed transition.

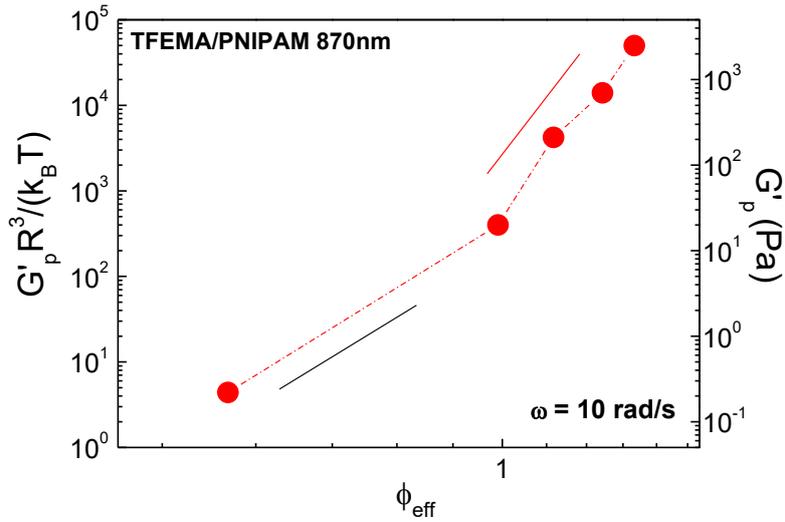

*Figure 3. Volume fraction dependence of the plateau modulus, $G'_P$, at rest for TFEMA/PNIPAM (870 nm) as a function of effective volume fraction. The solid lines indicate the two different power law dependencies at low and high effective volume fractions.*

Figure 4 includes comparisons for various rotational shear rates for the three different volume fractions measured. We observe that even at the lowest shear rate imposed ($\dot{\gamma}=0.01$ s$^{-1}$), the ODFS shows a discernible change from the DFS at rest, with the elastic modulus $G'_\perp$, decreasing and the viscous modulus $G''_\perp$, increasing over the whole frequency range. This finding is consistent with the idea that the rotational shear rate and consequent shear stress partially fluidizes the glassy sample.

Moreover, both $G'_\perp$, and $G''_\perp$, increase as the effective volume fraction increases for all the applied shear rates. At shear rate 0.01 s$^{-1}$, we detect the emergence of a crossover frequency, inside the experimental frequency window, which moves to higher frequencies with increasing shear rate. The crossover frequency provides a direct measure of the shear induced structural relaxation that is attributed to the microscopic yielding mechanism of cage breaking (Jacob et al. 2015). Above a certain shear rate, a second crossover frequency is observed at lower frequencies. These two crossover frequencies represent two structural relaxation timescales ($t_c=1/\omega_c$) that may be related with two different length-scales. As the shear rate is increased, we observe that both the high and the low crossover frequency $\omega_c$ move to higher values, denoting the speed-up of the internal dynamics. In the case of HS glasses, the high frequency crossover was related to the shear-induced out-of-cage displacements, while the low frequency crossover was attributed to the possible formation of larger length-scale structures or aggregates (Jacob et al. 2015) due to the interplay of shearing with lubrication hydrodynamics interactions (hydro-clusters) (Bossis and Brady 1985, Bender and Wagner 1996). Hybrid microgels though consist of a soft polymeric shell surrounding the hard core, defined by repulsive thermodynamic forces, which modify the lubrication interactions, minimizing the probability of hydro-clustering (Melrose and Ball 2004). Nevertheless, a similar phenomenon is detected here in jammed soft core-shell particles, suggesting either a universal phenomenon common to all these systems, besides the differences in the magnitude of hydrodynamic interactions or other surface interactions. However, before enquiring such mechanisms we need to exclude (or determine the extent of) any possible instrumental effects of the orthogonal probing at ultralow frequencies due to the yield stress nature of the measured samples.

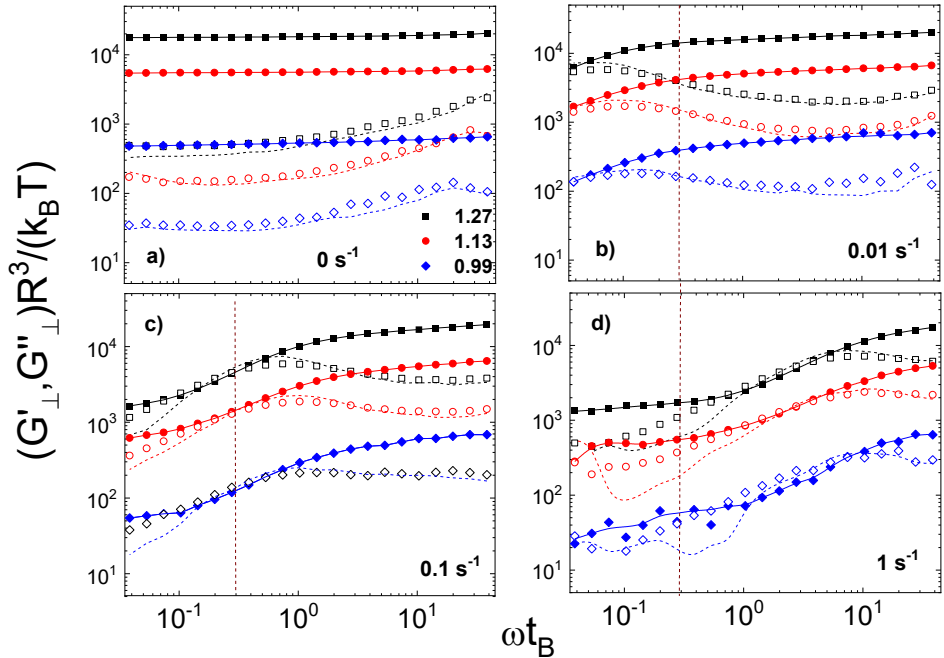

*Figure 4. Orthogonal dynamic frequency sweeps obtained for different rotational shear rates from a) at rest (0 $s^{-1}$) to f) 1 $s^{-1}$ at $\varphi_{eff}$ =0.99, 1.13 and 1.27 of TFEMA/PNIPAM (870 nm). $G'_\perp$, and $G''_\perp$ are*

*plotted as a function of the scaled frequency, $Pe_\omega = \omega t_B$. All measurements were conducted with the commercial ARES-G2, OSR geometry. Solid and dotted lines represent the Kramers - Kronig analyses*

*for G' and G'' respectively. Perpendicular dash lines indicate the lower $Pe_\omega$ limit for accurate OSR experiments as suggested by the open bottom Couette (OBC) measurements with the MCR-702, Anton-Paar rheometer. The unscaled data are shown in Figure S2 of the supplementary material.*

Hence, we use a new open bottom Couette (OBC) cell (figure A1) where pumping effects in highly viscous and yield stress samples are expected to be suppressed. Data from such OBC orthogonal set-up are shown in figure 5 for a series of applied shear rates at a high volume fraction ($\phi_{eff}$ =1.37) sample. The data of the loss modulus for the samples at rest and at small shear rates that are outside the operational window of the instrument as determined by the conservative data analysis applied, (as explained in appendix I), are not reported in figure 5. A more lenient (to that shown in Figure 5) data analysis of these measurements is shown in the supplementary Figure S3, with essentially the same message. Although the overall response of the material is confirmed, the curves lack a visible low-frequency crossover ($\omega_{cL}$). Even though only a limited frequency window is accessible due to resolution limitations, the behavior at frequencies lower than the high-frequency crossover ($\omega_{cH}$) suggests that the low-frequency crossover is not clearly detected, if not absent. Thus, it is reasonable to conclude that the low frequency response in our previous measurements (see figure 2b and figure 4 where an enhanced elasticity and a low-frequency G'-G'' cross-over is seen) is, to a certain extent, related with the residual pressure (pumping) effects present in the normal orthogonal (close bottom Couette) geometry and amplified by the elastoplastic (yield stress fluid) nature of the material. As these OSR experiments with the OBC cell indicate, the OSR data can be safely measured down to a low frequency limit of about $Pe_\omega$ = 0.3 (see figure 5), although this limit is expected to depend on the technical details of the geometry and transducer and hence on the rheometer used. We therefore indicate this low frequency limit in figure 4 noting that the orthogonal moduli data from sheared samples at lower frequencies might, to a certain extent, be affected by instrumental factors (of the normal OSR geometry).

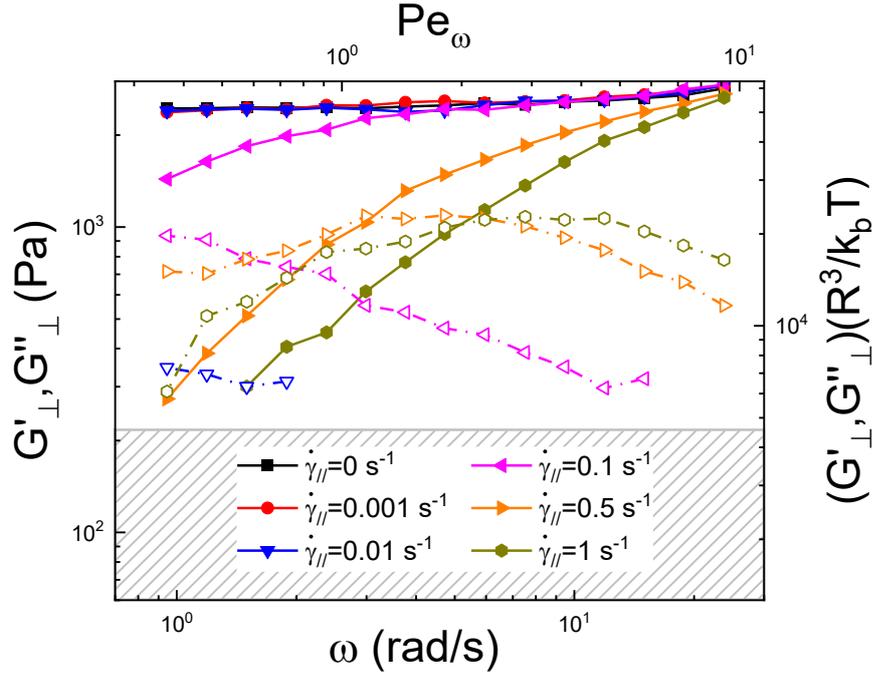

*Figure 5: Orthogonal dynamic frequency sweep obtained for different shear rates at $\varphi_{eff}=1.37$ of TFEMA/PNIPAM (870 nm) with the new OBC geometry. Solid lines with filled symbols represent the storage moduli, while the dash-dotted lines with empty symbols represent the loss moduli. The shaded grey region represents the lower limit of the operating window given by the minimum force measurable, all the datapoints with lower values are excluded from the plot.*

With the knowledge extracted from the new OBC measurements and the related analysis (in Appendix I) we can look again in the extended series of data presented in Figure 4 taken by the conventional (commercial) OSR geometry in the ARES-G2. We note that for the regime above $Pe_\omega = 0.3$ (identified as the low limit for the new OBC measurements) measurements of Figure 4 (c) and (d) (corresponding to high steady shear rates) still indicate the existence of a second low-frequency cross-over regime. These data fall within the instrument's (ARES-G2) operational window (see Appendix II) and can be directly compared with the OBC measurements. Since this second, low-frequency crossover is absent in the OBC geometry in this frequency range, we may reasonably assume that its origin relates to pressure effects (which are absent in the OBC geometry).

With possible instrumental artefacts being identified, we are now well placed to explore the validity of the Kramers-Kronig (KK) relationship in the soft colloidal glasses under steady shear probed by orthogonal superposition rheometry. The KK relationship, linking the elastic and viscous moduli, was thus utilized in one of its forms, i.e. $G''(\omega) \simeq \frac{\pi}{2}\left(\frac{dG'(u)}{d\ln u}\right)_{u=\omega}$, (Booij and Thoone 1982) to test its validity in the orthogonal viscoelastic moduli. To do so we first interpolated the $G'_\perp$ data with a second order polynomial function (shown as solid line among experimental data in figure 4). Subsequently, the interpolating data were used to calculate the $G''_\perp$ according to the above Kramers-Kronig relation, via the derivative of $G'_\perp$. The results of the model are presented in figure 4 as dashed lines. As expected for our soft glasses at rest the KK relationship is followed for the entire frequency range probed. For the samples under steady shear the resulting response of the orthogonal moduli verifies the validity of the KK relationship, especially at frequencies above the high frequency crossover, where the application of the relation yields an excellent agreement with experimental data, whereas below the low ω_c some deviations are detected. The validity of the Kramers-Kronig relation for the orthogonal in general denotes that the applied stationary shear flow does not induce considerable modification of the microgels' microstructure (Dhont and Wagner 2001). This confirms that while some distortion of the sample microstructure under shear occurs, it does not significantly differ from the microstructure at rest. Moreover, if we exclude the low frequency regime (Pe_ω < 0.3) where according to the discussion above we may experience instrumental effects, we do see that for higher frequencies deviations of the orthogonal viscoelastic data from KK are minimal. Note though that one could also use reversely this argument, arguing that that for these systems at least, the deviation from KK is a strong indication of the existence of instrumental artifacts in the measured data.

Figure 6 presents the relaxation times determined at the low and high frequency crossovers as a function of Pe. Note that although, as shown above the low-frequency cross-over is prone to instrumental effects we chose for completeness (and for comparison with earlier studies) to still present its response in the following. A direct comparison of microgel suspensions with HS glasses is possible normalizing the relaxation times with the Brownian time of each system in the dilute regime. The scaled short-time relaxation, deduced from the high frequency crossover, shows almost a linear decrease with Pe, similar to HS glasses ($t_{cH}/t_B \sim 1/Pe$) (Figure 6a). Increasing shear rate (above Pe≈1), we expect that the microstructure becomes more anisotropic, as seen in other studies in hard and soft particle glasses and jammed systems with an increasing radial distribution function g(r) in the compression axis and a decreasing in the extension axis (Seth et al. 2011, Koumakis et al. 2012a). For HS glasses this particle rearrangement is associated with the shear assisted particle cage escape and hence the reduction of the short-time relaxation (Jacob et al. 2015). Here, we observe that the short-time relaxation (corresponding to the high frequency cross-over) exhibits a non-monotonic dependence on particle volume fraction (Jacob et al. 2015) in contrast with HS glasses where the relaxation time was found to be independent of particle volume fraction. This non-monotonic dependence of short-time relaxation on particle volume fraction could be an indication of a transition from a glassy state to a jammed state, similarly to the change of the slope of the plateau modulus with particle effective volume fraction (see Figure 3). On the other hand, the long-time relaxation, (corresponding to low frequency crossover), exhibits also a power law decrease with Pe (Figure 6b), with a volume fraction dependent exponent of 0.8 or lower. Here data deduced in the low frequency regime where instrumental effects are definitely dominant (i.e. $Pe_\omega < 0.3$), are indicated with open symbols, while in view of the discussion above also for higher for higher frequencies they should be also considered with caution. We should point out that in the case of HS glasses, Brownian dynamics simulations, which do not take into account the hydrodynamic interactions (HI) between particles, exhibit only a high frequency crossover. Therefore, in a previous study the low frequency crossover detected only in experiments was associated to formation of hydroclusters which is more prominent at high shear rates (Jacob et al. 2015) but , was not directly verified, and in view of the instrumental effects identified here should now be reconsidered.

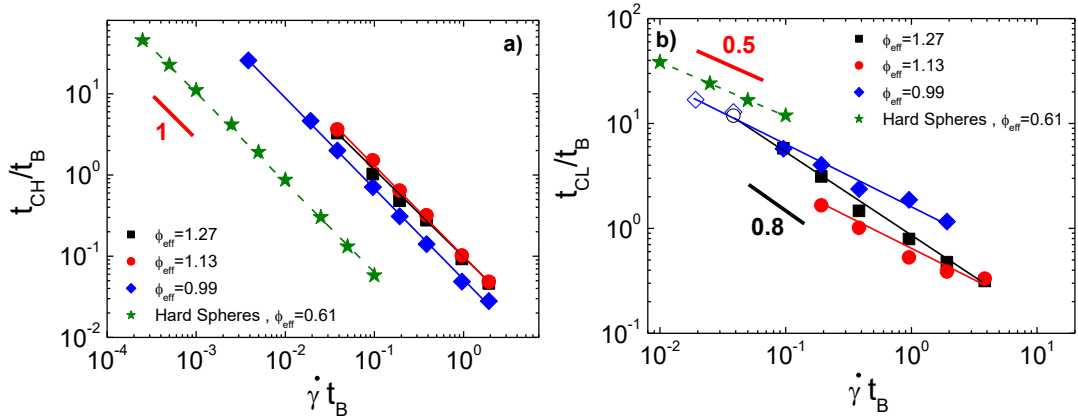

*Figure 6. Normalized relaxation times, ($t_{CH.L}/t_B$) deduced from (a) the high frequency cross-over frequency, $\omega_{CH}$ and (b) the low frequency cross-over frequency, $\omega_{CL}$, for various effective volume fractions as indicated. The results from experiments on hard sphere glasses at $\varphi =0.61$ are also shown for comparison. In (b) data for the low frequency cross-over time $t_{CL}/t_B$ deduced from figure 5 in the low frequency regime were OSR measurements definitely affected by instrumental errors are shown with open symbols*

The shear rate dependence of the crossover frequency was further used to investigate if a superposition of the viscoelastic spectra can be achieved (Jacob et al. 2015). Indeed, a scaling of the moduli as a function of frequency for the different steady state shear rates was used to superimpose all curves by shifting the experimental data in the x- (frequency) and y- (modulus) axes by factors, α and b, respectively. This was produced in a way that the high frequency crossover frequencies $\omega_c$ for all shear rates coincide. In the regime where the crossover frequency $\omega_c$ is not measurable, the shift is performed in a way that the rest of the viscoelastic spectra matches. The scaling results for the present core-shell microgel glasses at T=20°C are presented in Figure 7a, b and c for the three different effective volume fraction samples measured ϕ$_{eff}$=0.99, 1.13 and 1.27 respectively. Hence, a scaled mapping of the dynamics of the microgels through a shear rate-orthogonal frequency superposition (SROFS) is obtained. In all the measurements for the three different volume fractions, two distinct frequency regimes are detected which are separated by the crossover frequency $\omega_c$. In the high frequency regime, for $\omega>\omega_c$, which corresponds to short-time scales, the elastic moduli $G'_\perp$ superimpose for all shear rates, whereas the loss modulus $G''_\perp$ increases as shear rate is increased (see arrows in Figure 7). In contrast, for low frequencies, ω<ω$_c$ related to long-time scales, $G'_\perp$ exhibits a decrease with increasing shear rate, whereas $G''_\perp$ shows a rather good superposition for all shear rates explored. In comparison for the case of HS glasses, the deviation of $G''_\perp$ from a good scaling detected at high frequencies has been related to shear-induced reduction of short-time in-cage motions (Jacob et al. 2015). Shear deforms cages leading to some structural anisotropy with particles being accumulated more in the compression axis while are pushed out in the extension axis restricting the short-time in-cage particle diffusion (Koumakis et al. 2012a, Koumakis *et al.* 2016). In contrast, the deviation of $G'$ at low frequency which has been previously related to the formation of hydro-clusters in HS glasses, is probably not possible in soft core-shell particles, while in addition as discussed above should also be considered with caution as it is now shown to be affected by instrumental factors (for measurements with conventional OSR Couette geometries). Hence, in the regime where the instrumental factors are not important, the deviations (in the SROFS) for the present soft-hybrid microgels at low frequency can be related to local effects such as partial chain interpenetration or deformation in the outer shell of the particles which cannot be scaled by the same rate dependent factor as the overall structural relaxation of the high frequency crossover. Similar effects have been probed in concentrated star-like micelles by (Jacob *et al.* 2019) that had been attributed to long time relaxation processes under shear.

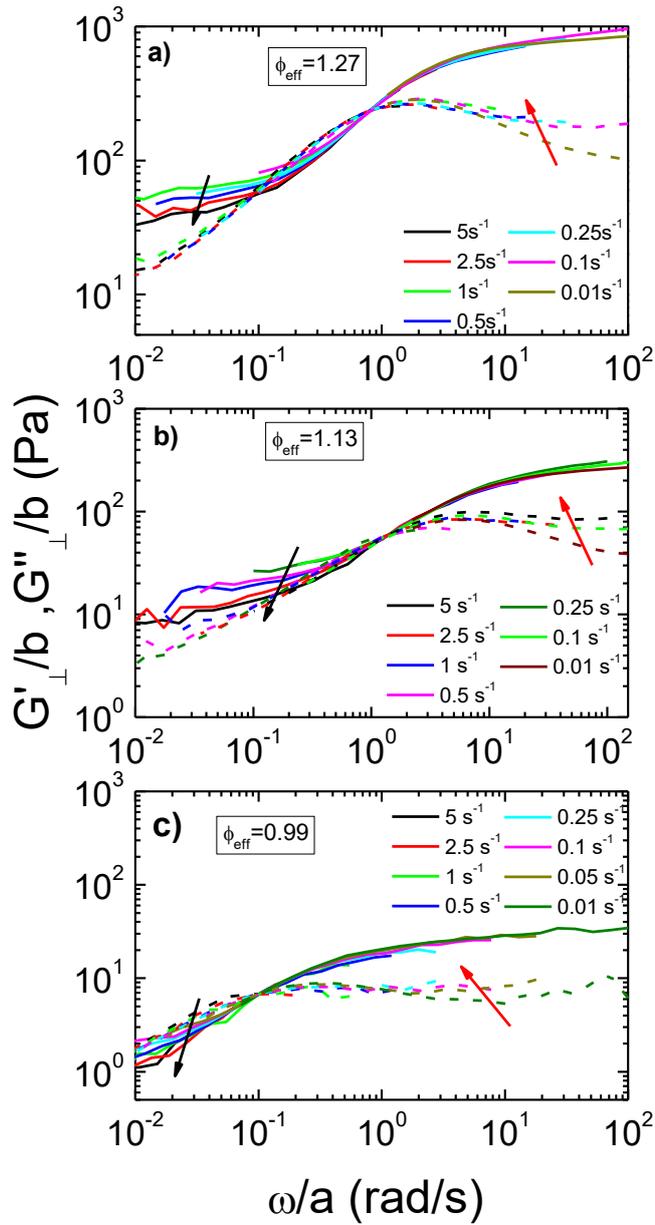

*Figure 7. Superposition of orthogonal dynamic frequency sweeps for microgels at various rotational shear rates for shear rates 0.01 $s^{-1}$ to 5 $s^{-1}$ for effective volume fractions (a) 0.99, (b) 1.13 and (c) 1.27. Arrows indicate in the low and high frequency regimes the increase of steady shear rate.*

The horizontal and vertical scaling factors used in obtaining the shear rate-orthogonal frequency superposition plots of Figure 7 are shown in Figure 8. The horizontal (frequency) shift factor reflects the shear rate dependence of the transition from the in-cage to out-of-cage motion and exhibits a linear dependence on the shear rate, similarly to that found in hard sphere glasses (Jacob et al. 2015). In comparison for ultrasoft star-like micelles the dependence of the shift factor on shear rate was found to be sublinear (with the power law exponent equal to 0.8) (Jacob et al. 2019). Moreover, the frequency shift factor of the present hybrid core-shell microgels also exhibits a non-monotonic change with the effective volume fraction, similarly with the behavior of the $\omega_c$. As mentioned above this non-monotonic dependence could be signaling the transition from a glassy to the jammed state, as it is evident by a significant change in the elastic modulus with particle volume fraction (Figure 3). To the contrary, repulsive hard sphere (HS) glasses present a volume fraction independent shear rate dependence of the shift factor α (Jacob et al. 2015).

For HS glasses, the vertical shift factor b reflects the effect of shear on the local free volume inside the cage and is directly connected to entropic elasticity of the hard sphere glass. Increasing the shear rate, the short-time in-cage motion of the particles is slowed down as particles are constricted in a deformed cage. When converted to viscoelastic moduli, using the generalized Stokes-Einstein relation (Mason 2000), the decreased in-cage motion corresponds to stronger elasticity and thus $G'$ is higher. Furthermore, for lower effective volume fractions, a loosely packed cage is able to deform more under shear before breaking due to higher in-cage free volume, which is also reflected to the shear rate dependence of the modulus shift factor (Jacob et al. 2015). To the contrary, at higher volume fractions, a tightly packed cage cannot deform much under shear due to smaller in-cage free volume with the modulus shift factor being almost shear rate independent and close to 1 (Jacob et al. 2015). The behavior of vertical shift factor for the present hybrid core-shell microgels is analogous to that of the HSs, in the sense that it increases stronger with shear rate at lower volume fractions while at higher volume fractions in the jammed state it is close to 1 and exhibits a weak shear rate dependence. It worth noting that for the ultrasoft star-like micelles (Jacob et al. 2019) this shift factor is essentially shear rate independent and equal to 1. This behavior is possibly related to their ability to partially interpenetrate each other and easily deform when they are jammed.

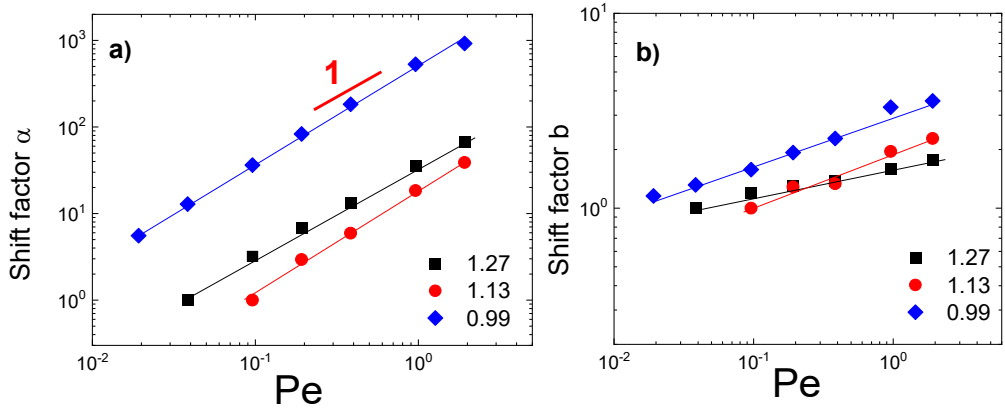

*Figure 8. a) The horizontal (frequency) shift factor α and b) the vertical (modulus) shift factor b used for SROFS analysis for samples with $\varphi_{eff}$ = 0.99, 1.13 and 1.27.*

## IV. Conclusions:

OSR was employed to study shear-induced microstructural changes of soft hybrid core-shell microgels. The orthogonal superposition moduli, extracted during steady state flow at various shear rates, provide information about the mechanism of flow under shear at different effective volume fractions well above glass transition. At rest, hybrid microgels exhibit a typical behavior of colloidal glass with G'>G'' in a wide frequency range. Under steady shear flow, for all the probed volume fractions, two crossover frequencies were observed, which move monotonically to higher values increasing the applied shear rate and represent two structural relaxation timescales ($t_c=1/\omega_c$) with the high frequency crossover related to the shear-induced out-of-cage displacements.

Superposition of the orthogonal viscoelastic spectrum under different shear rates was realized by using shift factors α and b. Both shift factors exhibit a power law dependence on the shear rate. The frequency shift factor, which is directly connected to $\omega_c$, exhibits an almost linear dependence that is similar to that in repulsive hard spheres, despite the soft character of the hybrid core-shell microgels. This can be attributed to the crosslinked shell that exists in the outer part of the particles, which prohibits full chain interpenetration allowing only a partial one, and its ability to deform. Contrary to HS behaviour, the frequency shift factor for microgels presents volume fraction dependence, which may be due to the transition from a glassy to a jammed state. The modulus shift factor, which is connected to the elasticity under shear and expresses the in-cage free volume, shows a shear rate dependence both at low and high volume fractions. This behaviour differs for HS, where at lower volume fractions, shear affects the vertical shift factor, due to increased free volume inside a loosely packed cage but at higher volume fractions, a tightly packed cage cannot deform under shear before breaking, rendering the factor b shear independent.

Moreover, we interrogated the low frequency response, via the use of a new OSR Couette cell which eliminates pumping effects in OSR measurements of yield stress fluids and reveals the true low frequency response and the OSR low frequency limits. Finally, we applied the Kramers-Kronig relationship on the orthogonal superposition viscoelastic spectra from our sheared soft colloidal yield stress states and revealed its applicability regime. Its validity in a wider range of frequencies suggests the absence of strong new shear induced structural changes, while KK deviations at low frequencies, in agreement with the new OBC OSR measurements, indicate the regime where instrumental artefacts in OSR might be present. The KK validity approach appears thus appealing to be implemented in systems with demonstrated strong shear induced structural anisotropy, such as fractal colloidal gels, or when strong structural changes are absent as a mean to distinguish instrumental effects in OSR experiments.

**Acknowledgements:**

The authors acknowledge funding from the EU program EUSMI. Anton Paar (Dr. J. Lauger) is gratefully acknowledged for the use of the OSP prototype. GP acknowledges funding from Twinning project FORGREENSOFT (Number: 101078989 under HORIZON WIDERA-2021-ACCESS-03). JV acknowledges the Swiss National Science Foundation, SNSF Grant No. 200020-192336

**Appendix I:**

The data presented here are the first published data acquired with the open bottom Couette (OBC) and it is important to clarify the refinements applied to them. Instead, the linear motor prototype was already tested using a standard orthogonal couette in Pagani et al. (2024). A schematic design of the new OBC couette is presented in Figure A1.

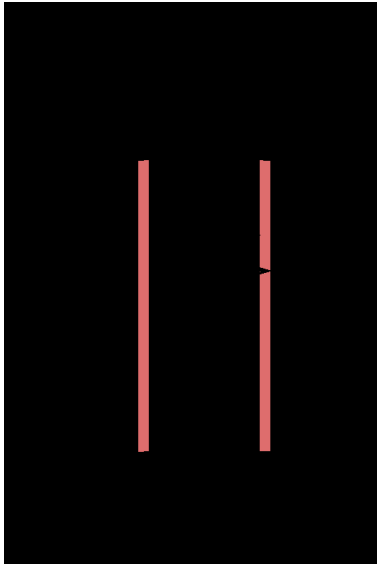

*Figure A1. Schematic representation of the new OBC geometry. The inner bob controls the rotational motion, while the outer cylinder control the linear oscillation. The outer cylinder is connected with the geometry shaft leaving an empty space at the bottom. In pink colour is highlighted the volume occupied by the sample.*

The instrument (MCR702, Anton-Paar) operates and records the raw variables (force, speed and displacement) and these must be converted manually to relevant quantities. The moduli were calculated with the formulas proposed by Vermant et al. (1997). The formulas are reported in equations (1) and the geometrical coefficient A and β are adapted for the used geometry.

1.1. $G'_\perp = \frac{1}{A}\left(\frac{F}{s}\cos\delta - K + (m + \beta A\rho)\omega^2\right)$

1.2. $G''_\perp = \frac{1}{A}\left(\frac{F}{s}\sin\delta + \zeta\omega\right)$

1.3. $A = \frac{2\pi L}{\ln r}$

1.4. $\beta = \frac{R_O^2}{2}\left(\frac{r^2-1}{2r\ln r} - 1 + \ln r\right)$

The previous formulas correct the measured force $(F)$, normalized by the applied displacement $(s)$, from the instrumental artifacts: the elastic $(K/A)$, the friction $(\zeta/A \; \omega = \zeta^* \omega)$, and the inertia $(m/A \; \omega^2)$ of the motor and the sample inertia $(\beta A \rho \; \omega^2)$. The effect of sample and motor inertia can be combined in an inertia modulus that represent the impact of the system inertia on the measurement, $G'_I = (m + \beta A \rho)/A \; \omega^2 = I \; \omega^2$. The phase angle $(\delta)$ is measured directly by the MCR702. The coefficient A and β are function only of the dimensions of the geometry represented in Figure A1: the measuring length $L$, the bob radius $R_o$, the cup radius $R_o$, and their ratio $r$ $(r = R_i/R_o)$.

The limiting angular frequencies $(\omega_{min}, \omega_{max})$ and the minimum modulus $(G^*_{min})$ are defined by instrument limitations and quantified by observing the raw force measured. To let the reader observe first-hand, the raw measured force as function of frequency is presented in Figure A2 for the system under study (Figure A2).

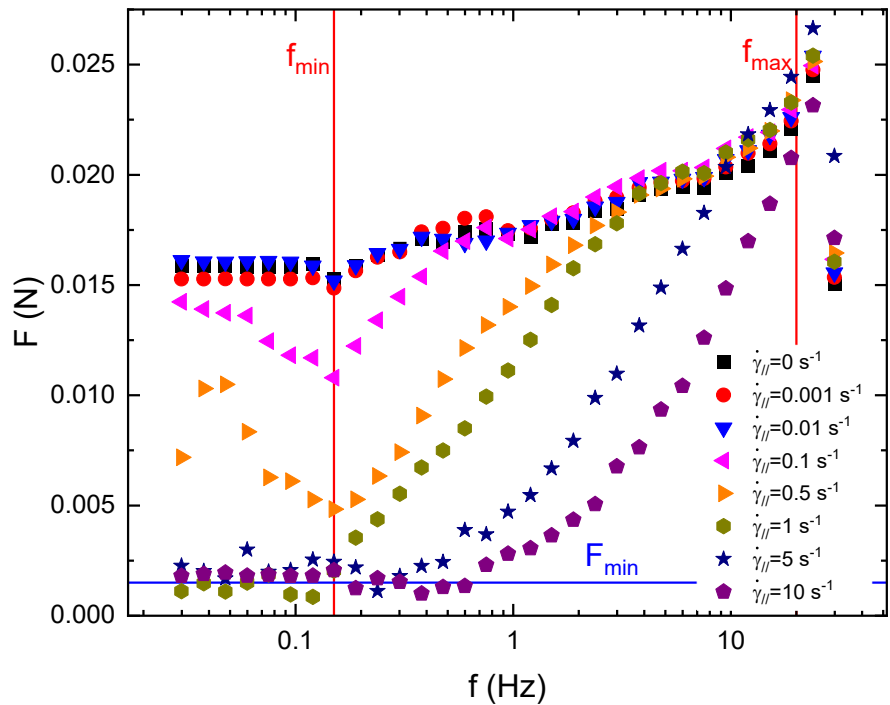

*Figure A2. Force as a function of frequency as measured by the instrument (MCR702, Anton-Paar) at $\varphi_{eff}=1.37$ of TFEMA/PNIPAM (870 nm). The frequency range is clearly identifiable by the sudden deviation of the force measurements. While the minimum force is particularly recognizable in the measurements at $5s^{-1}$ and $10s^{-1}$.*

In Figure A3 the complete range of moduli collected for the particles studied with this cell is represented with the instrument limitations and the applied corrections. A conservative approach was used in data refinement. All the data points were discarded if their value was lower than the complex modulus derived from the minimal force, i.e. no phase angle decomposition was considered.

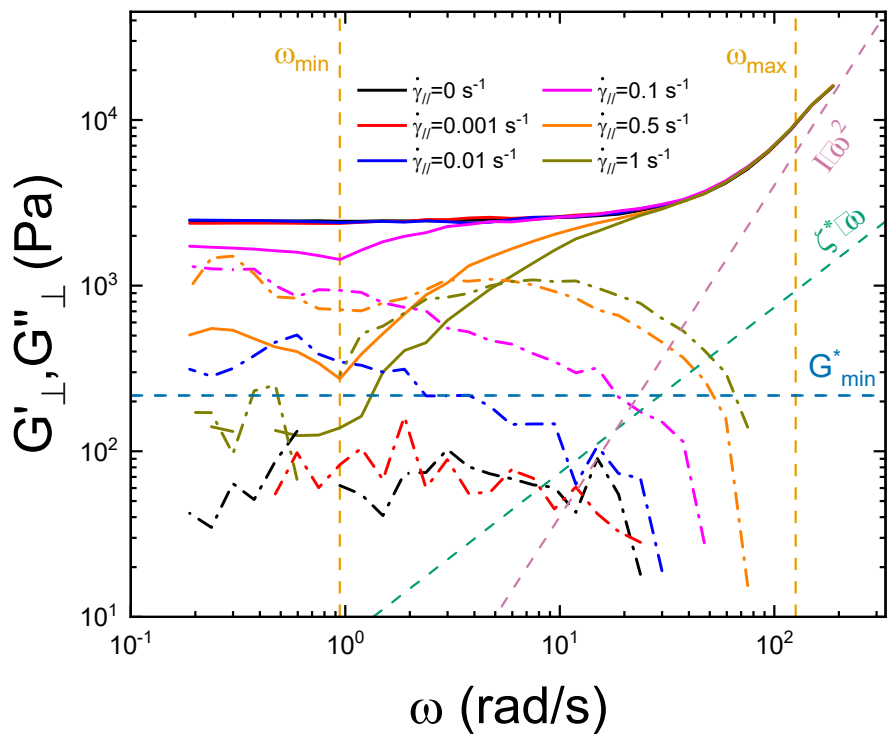

*Figure A3. Orthogonal dynamic frequency sweep obtained for different shear rates at $\varphi_{eff}=1.37$ of TFEMA/PNIPAM (870 nm), with MCR 702 (Anton-Paar). Solid lines represent the storage moduli, while the dash-dotted lines represent the loss moduli. The instrumental limitations, the friction correction, and the inertia correction are presented with dashed lines.*

*The inertia correction is the most impactful correction applied to the data. The percentage of response given by inertia is plotted in Figure A4. The final values reported in Figure 4 are obtained excluding data points where inertia contributes to more than 10% of the final value.*

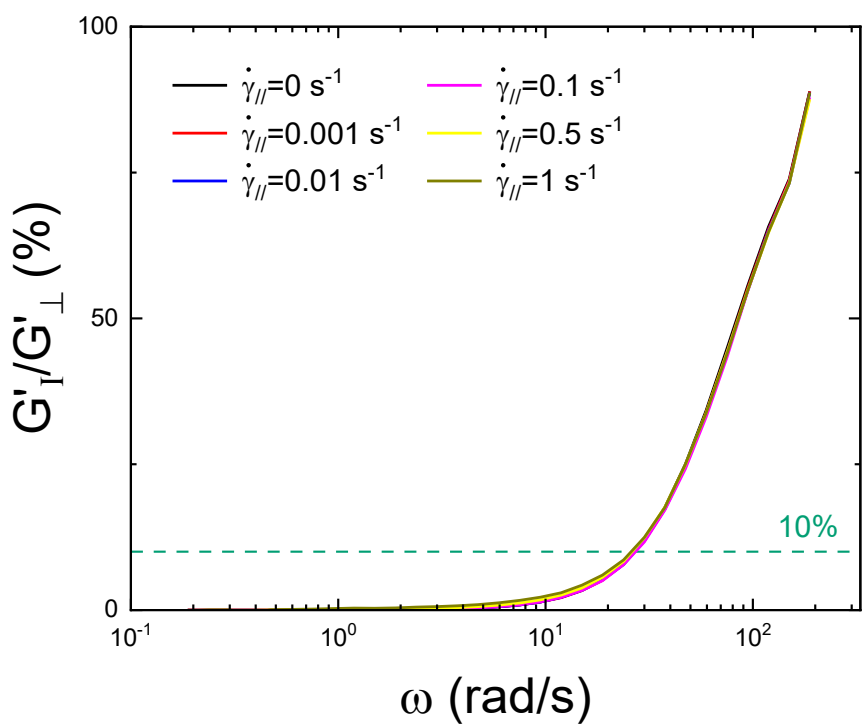

*Figure A4. Inertia contribution (MCR702, Anton-Paar) of the orthogonal dynamic storage modulus obtained for different shear rates at $\varphi_{eff}$ =1.37 of TFEMA/PNIPAM (870 nm). The dashed line represents the upper limit of the inertia acceptable in the measurements.*

Appendix II:

Below we present analysis of the ARES-G2 OSR measurements limits along the same lines as that discussed in Appendix I for the Anton-Paar MCR702.

The minimum and maximum frequencies as the minimum torque are the most impactful instrumental limits for the ARES-G2 rheometer. These are obtained by the specification given by the manufacturer. In FigureA5 the operational window is overlaid to the complex modulus ($G^*$) for the particles with $\phi_{eff}$=1.27. The operational window for the remaining particle concentrations is presented in the supplementary materials.

Based on the limits identified here in a stringent way (as for the new OBC cell with the MCR702, Anton-Paar measurements), the ODFS data highlight the uncertainty in the low frequency measurements due to instrumental effects.

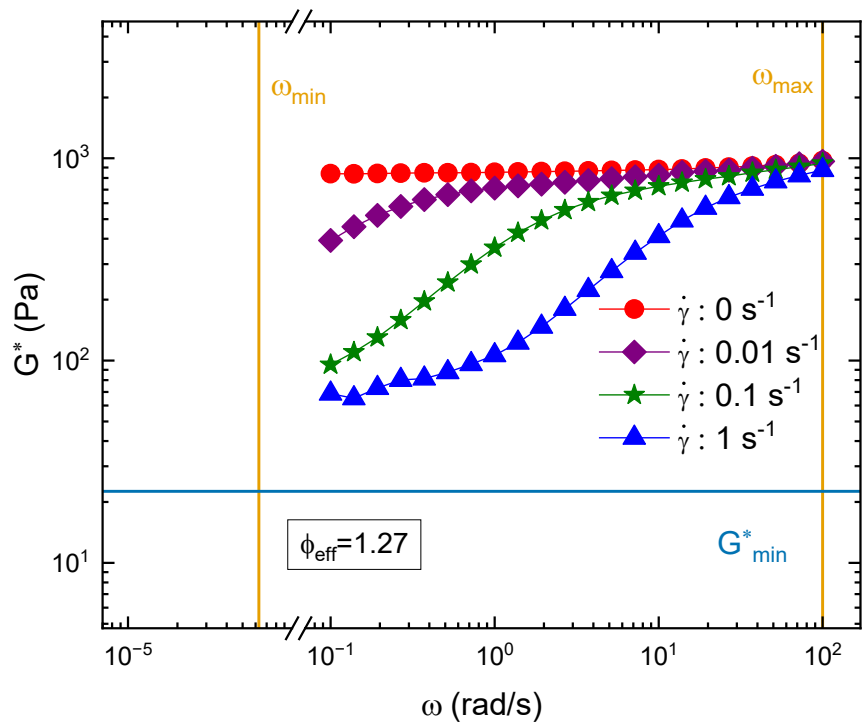

*Figure A5. Complex modulus as function of frequency operational window of the ARES-G2 rheometer overlayed to the complex modulus as a function of frequency for the sample at $\varphi_{eff} = 1.27$*

**Supplementary Material for**

# Orthogonal Superposition Rheometry of soft core-shell microgels

*Panagiota Bogri[1,2], Gabriele Pagani[3], Jan Vermant[3], Joris Sprakel[4] and George Petekidis[1,2]*

To allow a direct comparison with other experimental studies, data of the flow curves presented in Figure 1, are also shown in an unscaled form (stress and viscosity versus shear rate) in Figure S1. Similarly, unscaled orthogonal superposition data of G' and G'' corresponding to figures 4 of the main paper are depicted in Figures S2. In S2 we also apply the most stringent data refinement, as explained in the appendix. In Figure S3 the operational window for the ARES-G2 is overlaid to the data from the samples at $\phi_{eff}$ = 1.13 and $\phi_{eff}$ = 0.99, as done in Figure A5 for the sample at $\phi_{eff}$=1.27.

Figure S4 shows a more lenient data analysis, compared to the one presented in Figure 5, approach of measurements with the new OBC geometry in the most concentrated sample. In this analysis the effect of the minimum torque was decomposed on the storage and loss components according to the phase angle measured. Finally, in Figure S4 we show the OSR data at different $\phi_{eff}$ from measurements with the old and new OBC geometry at the same high steady shear rate. Data are scaled by the G' at the high frequency cross-over frequency and shown as a function of $\omega/\omega_{c,H}$. Figure S3 and S4 demonstrate how the pressure field notably affects the material behavior at low frequencies, particularly when compared to the new OBC measurements.

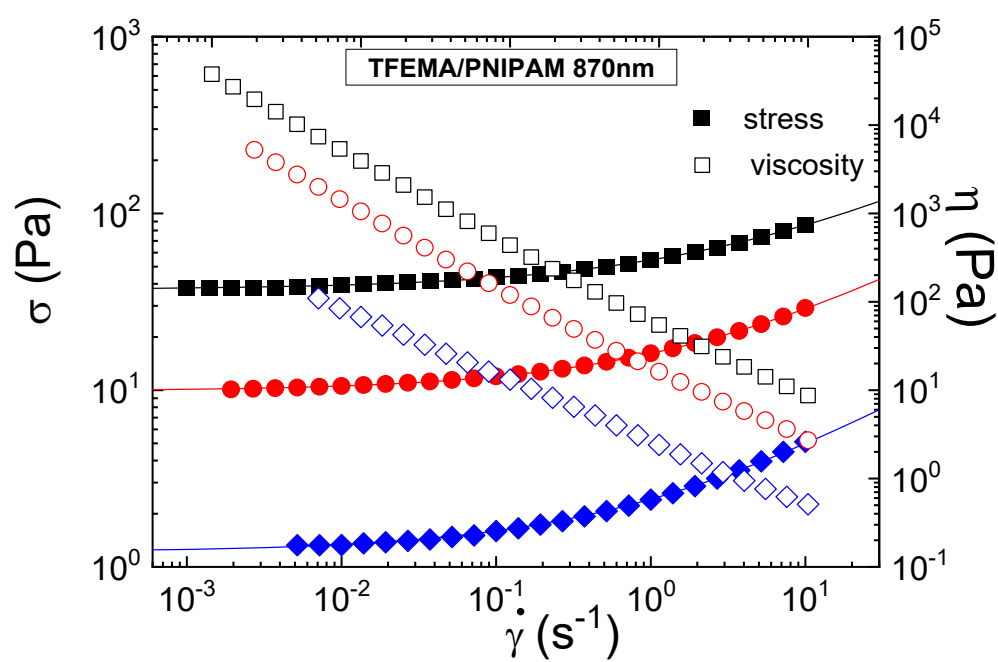

*Figure S1: Stress and viscosity under steady state shear as a function of shear rate for various volume fractions at T=20 °C.*

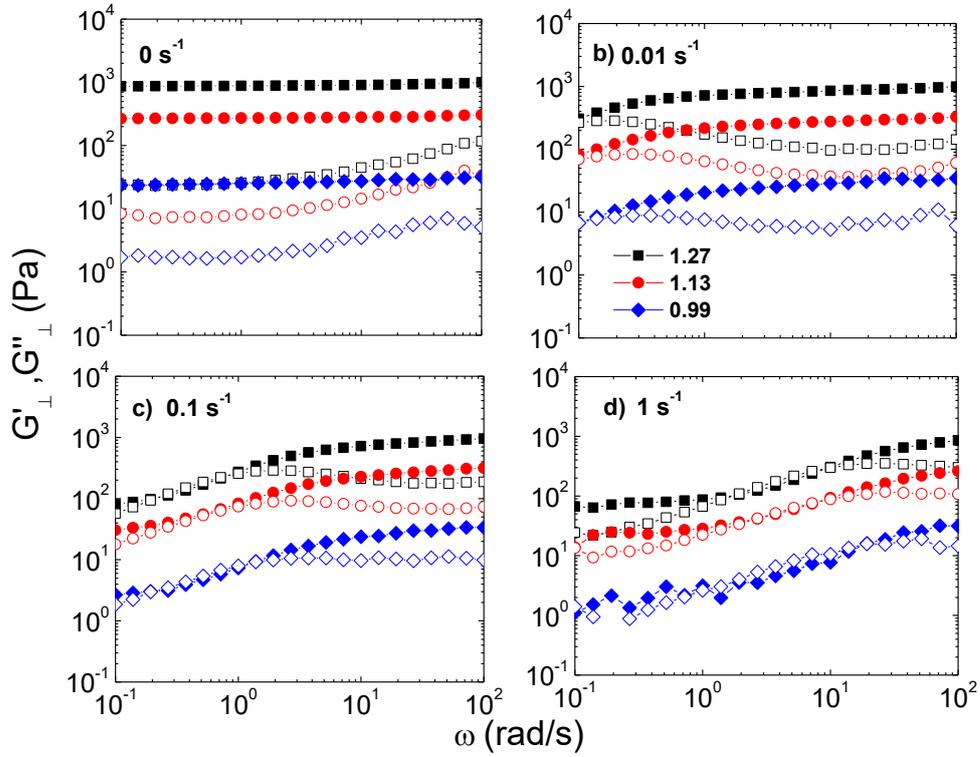

*Figure S2: Experimental data from Orthogonal dynamic frequency sweeps obtained for different rotational shear rates from a) at rest (0 s⁻¹) to f) 1 s⁻¹ at $\varphi_{eff}$ =0.99, 1.13 and 1.27 of TFEMA/PNIPAM (870 nm). $G'_\perp$, and $G''_\perp$ are plotted as a function of the frequency, $\omega$. Data correspond to those presented in Figure 4 as a function of $Pe_\omega = \omega t_B$*

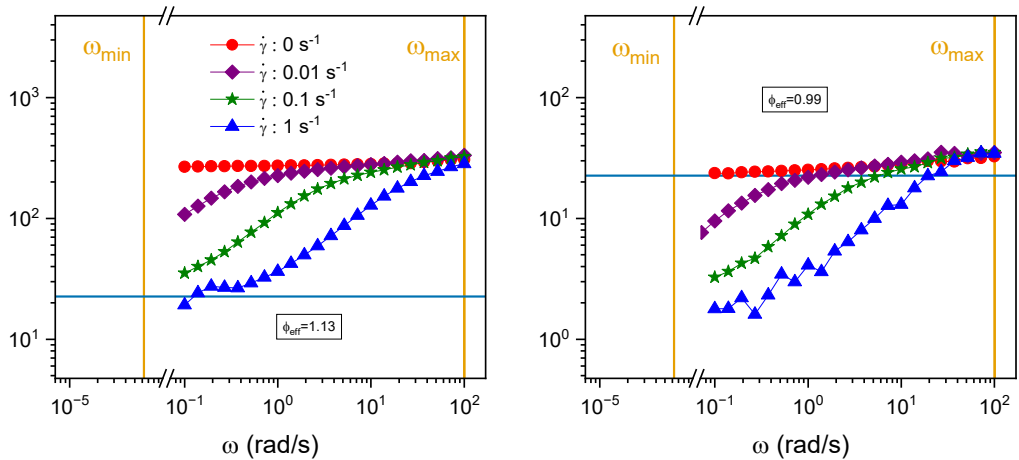

*Figure S3: Complex modulus as function of frequency. Operational window of the ARES-G2 rheometer overlayed to the complex modulus as a function of frequency for the sample at $\varphi_{eff} = 1.13$ and $\varphi_{eff} = 0.99$.*

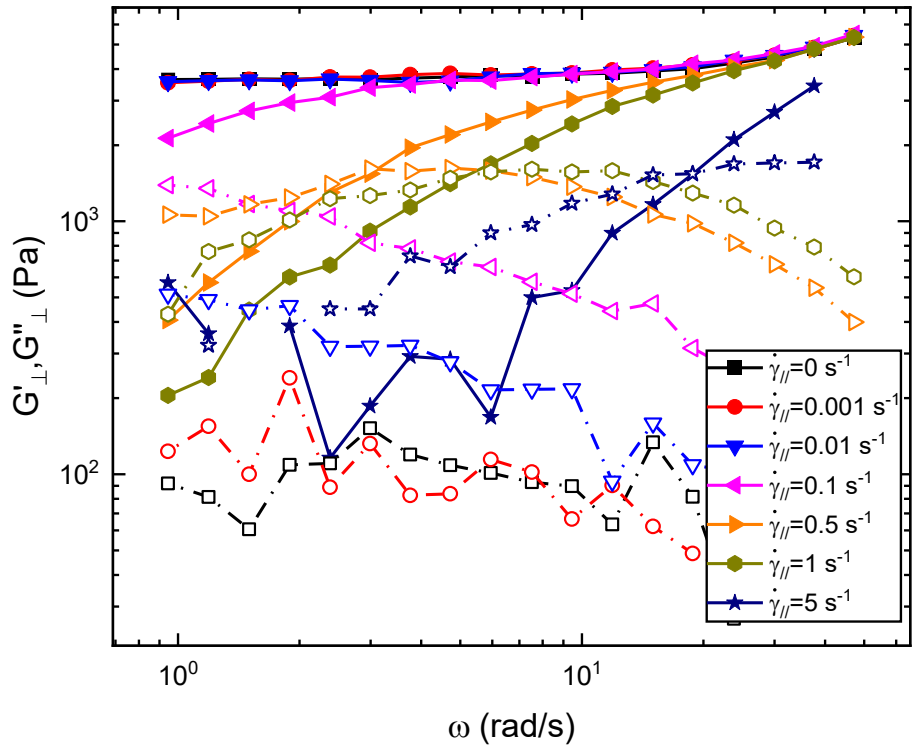

*Figure S4: Orthogonal dynamic frequency sweep obtained for different shear rates at φ$_{eff}$=1.37 with the new OBC geometry. Solid lines with filled symbols represent the storage moduli, while the dash-dotted lines with empty symbols represent the loss moduli. This plot provides a more lenient data analysis approach, compared to what presented in Figure 5*

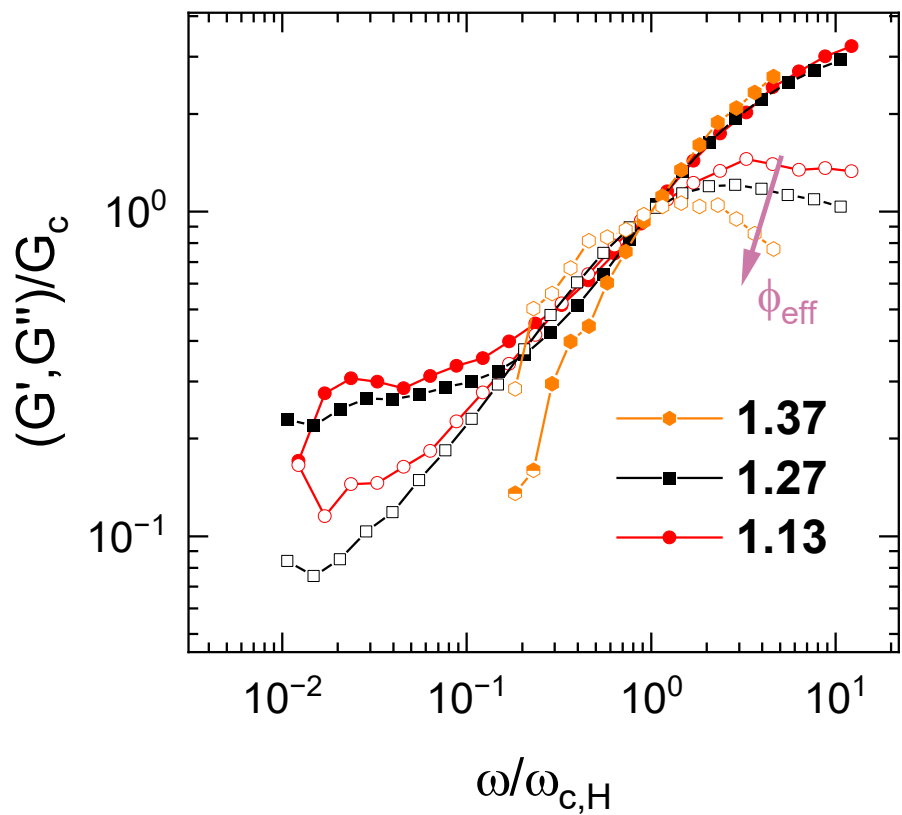

*Figure S5: OSR data at different $\varphi_{eff}$ at the same high steady shear rate (1 s$^{-1}$). Data are scaled by the G' at the high frequency cross-over frequency and shown as a function of $\omega/\omega_{c,H}$. The two data points partially solid are found with the lenient data refinement and are here included to extend the storage and the loss modulus for the same frequency extension.*